\documentclass[a4paper,margin=1.5cm]{article}

\usepackage{WEJCF}

\usepackage[utf8]{inputenc} % allow utf-8 input
\usepackage[T1]{fontenc}    % use 8-bit T1 fonts
\usepackage{siunitx}
\usepackage{hyperref}       % hyperlinks
\usepackage{url}            % simple URL typesetting
\usepackage{booktabs}       % professional-quality tables
\usepackage{lipsum}
\usepackage{ucs}
\usepackage{amsmath}
\usepackage{amsfonts}
\usepackage{amssymb}
\usepackage{graphicx}
\usepackage{setspace}
\usepackage{rotating}
\usepackage{multirow}
\usepackage{mathrsfs}
\usepackage{subfig}
\usepackage{bigstrut}
\usepackage{pdfpages}
\begin{document}
% CHANGE THE NAME OF THIS FILE ACCORDING YOUR ID IN INDICO (E.G. content_07.tex)

%your talk title
\title{Jet substructure measurements elucidating partonic evolution in $p$+$p$ collisions at RHIC}

%your name
\author{ Monika Robotková\\ \textit{on behalf of the STAR Collaboration} }
\maketitle
% Abstract (Do not insert blank lines, i.e. \\) 
\abstract{Jets are multiscale objects that connect partons to hadrons, making jet substructure measurements crucial for probing both perturbative and non-perturbative processes in QCD. At STAR, a variety of jet substructure observables, such as SoftDrop groomed splittings and N-Point Energy Correlators (ENC), provide insights into parton evolution and hadronization mechanisms. SoftDrop-groomed observables and ENCs both connect measurement to fundamental QCD at the parton level, allowing for comparisons to first-principles theoretical calculations. Additionally, by also including charge information, as in the charge-weighted ENC, details about the hadronization mechanism can be obtained.
In these proceedings, we present preliminary results on measurements of SoftDrop observables and ENCs across different jet momenta and radii in $p$+$p$ collisions at $\sqrt{s}$~=~200~GeV using STAR data.}

% start writing your roceeding here:
\section{Introduction}
Jets are collimated sprays of hadrons resulting from the fragmentation and hadronization of high-energy partons produced in hard QCD scatterings. The internal structure of jets, or jet substructure, provides insight into both perturbative and non-perturbative QCD dynamics. Studies of jet substructure in proton-proton ($p$+$p$) collisions serve as a baseline for understanding medium-induced modifications in heavy-ion collisions, as well as testing QCD-based models and Monte Carlo generators.

Jet substructure can be probed using several techniques. One common method is jet grooming, which removes soft wide-angle radiation and reveals the hard core of the jet. In particular, the SoftDrop \cite{SoftDrop} algorithm removes soft radiation based on condition:

\begin{equation} 
\frac{\min(p_{\text{T,1}},p_{\text{T,2}})}{p_{\text{T,1}}+p_{\text{T,2}}}>z_{\text{cut}}\left(\frac{\Delta R_{12}}{R}\right)^{\beta},
\label{soft_drop}
\end{equation}

\noindent where $p_{\text{T},i}$ is the transverse momentum of the corresponding subjet, $R$ is the resolution parameter of the jet, and $\Delta R_{\text{12}}$ is the distance between the two subjets. There are two parameter, $\beta$ and $z_{\text{cut}}$. The angular exponent $\beta$, controlling how the grooming depends on the angle between subjets. A threshold parameter $z_{\text{cut}}$ sets the minimal momentum sharing required for a branching to be considered hard and not groomed away. While the CollinearDrop \cite{CollinearDrop} method targets soft radiation by comparing observables under different grooming settings. These tools allow us to investigate the jet shower evolution at different stages.

In addition to grooming-based observables, energy-energy correlators (EECs) \cite{EEC} offer a complementary approach by studying angular correlations among final-state particles within a jet. These observables enable separation of perturbative and non-perturbative regimes and have the advantage of not relying on jet reclustering procedures. The projected $N$-point correlators, including the two-point (EEC) and three-point (E3C) versions, provide powerful handles on the multi-prong structure of jets and their scale evolution.

\section{Experimental Setup and Data Sample}
The analysis is based on $p$+$p$ collision data at $\sqrt{s} = 200$ GeV collected by the STAR experiment \cite{STAR_overview} at RHIC in 2012. Jets are reconstructed from charged particles measured in the Time Projection Chamber (TPC) \cite{TPC} and neutral energy deposits from the Barrel Electromagnetic Calorimeter (BEMC) \cite{BEMC}. Jet reconstruction employs the anti-$k_{\text{T}}$ algorithm \cite{anti_kT} with radius parameters $R=0.4$ and $R=0.6$.

Charged particles are selected with $0.2 < p_{\text{T}} < 30$ GeV/$c$, while neutral towers must satisfy $0.2 < E_{\text{T}} < 30$ GeV. Events are triggered using the jet patch (JP) trigger implemented in the BEMC, which requires the uncorrected summed patch ADC value to exceed $\sum E_{\text{T}} > 7.3$ GeV within one of eighteen partially overlapping regions of size 1.0~$\times$~1.0 in $(\eta,\phi)$, each defined by a grouping of calorimeter towers. The primary vertex is required to be reconstructed within $|v_{\text{z}}| < 30$ cm.

The SoftDrop grooming procedure is applied using reclustering with the Cambridge/Aachen (C/A) algorithm \cite{CA_2}. The SoftDrop parameters are chosen as $(z_{\text{cut}}, \beta) = (0.1, 0)$. For CollinearDrop, two grooming settings are compared: $(z_{\text{cut},1}, \beta_1) = (0, 0)$ and $(z_{\text{cut},2}, \beta_2) = (0.1, 0)$.

\section{Jet Substructure Observables}
We investigate several jet substructure observables derived from grooming techniques and constituent-level correlations. One of the fundamental observables is the shared momentum fraction $z_{\text{g}}$, defined as
\begin{equation}
z_g = \frac{\min(p_{\text{T,1}}, p_{\text{T,2}})}{p_{\text{T,1}} + p_{\text{T,2}}},
\end{equation}
which characterizes the momentum balance of the first hard splitting in the SoftDrop-declustered jet. Complementary to this, the groomed radius $R_\text{g}$ represents the angular distance $\Delta R_{12}$ between the two subjets that satisfy the SoftDrop condition, giving insight into the spatial extent of the splitting.

The splitting scale $k_\text{T}$ is defined as
\begin{equation}
k_\text{T} = z_\text{g} \cdot p_{\text{T,jet}} \cdot \sin R_\text{g},
\end{equation}
which combines the momentum and angular information into a single variable, probing the phase space of the Lund plane.

Jet mass $M$ and groomed jet mass $M_{\text{g}}$ are computed from the four-momentum of the jet constituents before and after grooming, respectively:
\begin{equation}
M = \sqrt{E^2 - {|\vec{p}|}^2}.
\end{equation}
Their difference normalized to the original mass defines the CollinearDrop mass ratio:
\begin{equation}
\frac{\Delta M}{M} = \frac{M - M_\text{g}}{M},
\end{equation}
which quantifies the amount of soft radiation removed by the grooming procedure. The jet mass serves as a sensitive probe of the internal structure of jets, reflecting the virtuality of the initiating parton and capturing both perturbative radiation and non-perturbative effects such as hadronization.

Beyond grooming-based observables, we utilize energy-energy correlators to study the angular correlations of final state charged particles. The two-point projected energy-energy correlator (EEC) is defined as
\begin{equation}
\text{Normalized EEC} = \frac{1}{\sum_{\text{jets}} \sum_{i \neq j} \frac{E_i E_j}{p_{T,\text{jet}}^2}} 
\cdot \frac{\text{d} \left( \sum_{\text{jets}} \sum_{i \neq j} \frac{E_i E_j}{p_{T,\text{jet}}^2} \right)}{\text{d}(R_L)}
\end{equation}
where $E_{i}$, $E_{j}$ are the energies of the two constituents and $R_{L}$ is the angular separation between them. This correlator captures the structure of jet radiation without requiring a clustering algorithm.

To explore more complex topologies such as three-prong splittings, we include the three-point correlator (E3C), defined as
\begin{equation}
\text{Normalized E3C} = \frac{1}{\sum_{\text{jets}} \sum_{i \neq j} \frac{E_i E_j E_k}{p_{T,\text{jet}}^3}} 
\cdot \frac{\text{d} \left( \sum_{\text{jets}} \sum_{i \neq j} \frac{E_i E_j E_k}{p_{T,\text{jet}}^3} \right)}{\text{d}(R_L)}
\end{equation}
These observables allow for a comprehensive, grooming-independent exploration of radiation patterns and color coherence within jets. They also allow for isolation of perturbative effects.

Altogether, the combination of SoftDrop, CollinearDrop, and energy correlators enables a multifaceted investigation of jet internal dynamics across different kinematic regimes and stages of evolution.

\section{Results}
\begin{figure}[htb]
\center
\includegraphics[width=0.5\linewidth]{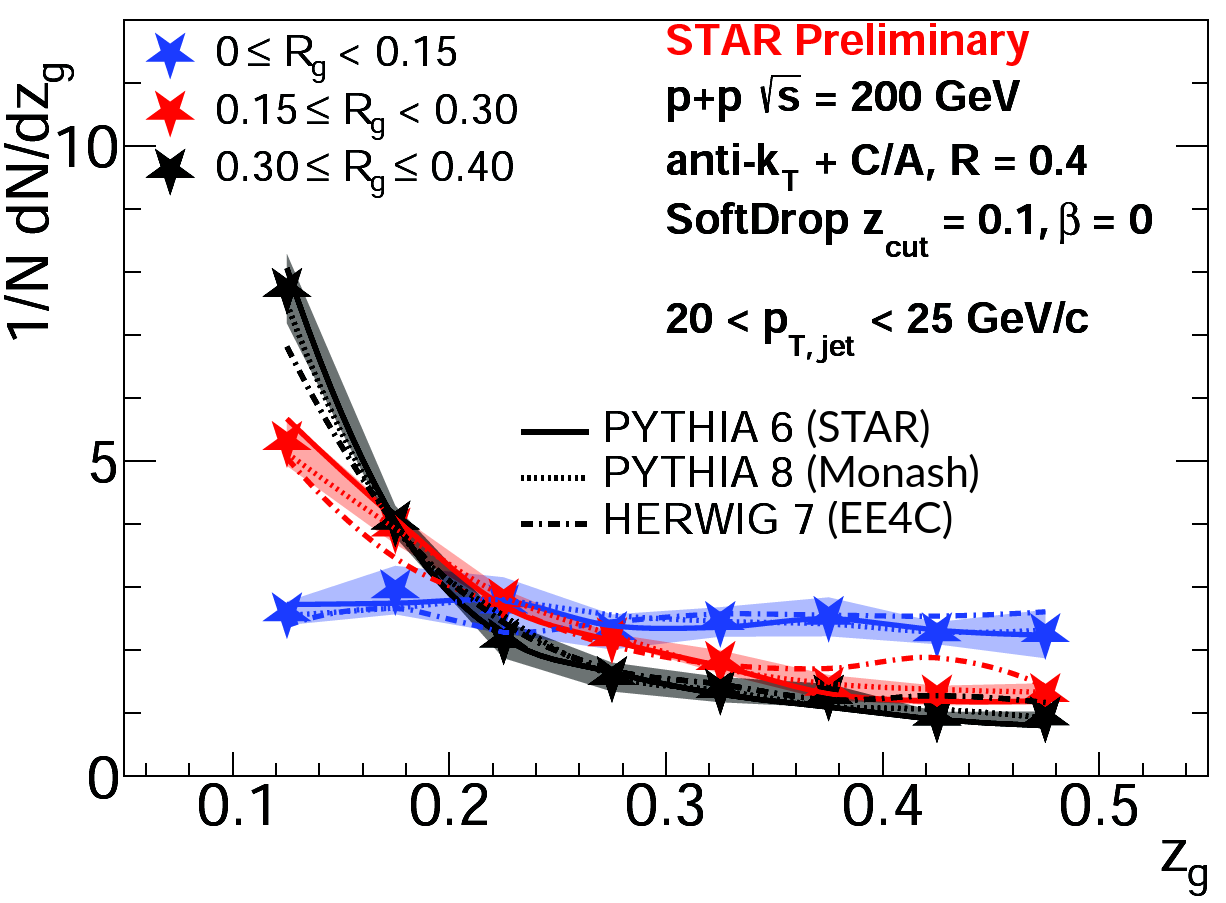}
\caption{Correlation between $z_{\text{g}}$ and $R_{\text{g}}$ at the first split for jets with $R$ = 0.4 in $p$+$p$ collisions at $\sqrt{s}$~=~200~GeV. \label{zg_rg_1}}
\end{figure}

\begin{figure}[htb]
\center
\includegraphics[width=0.55\linewidth]{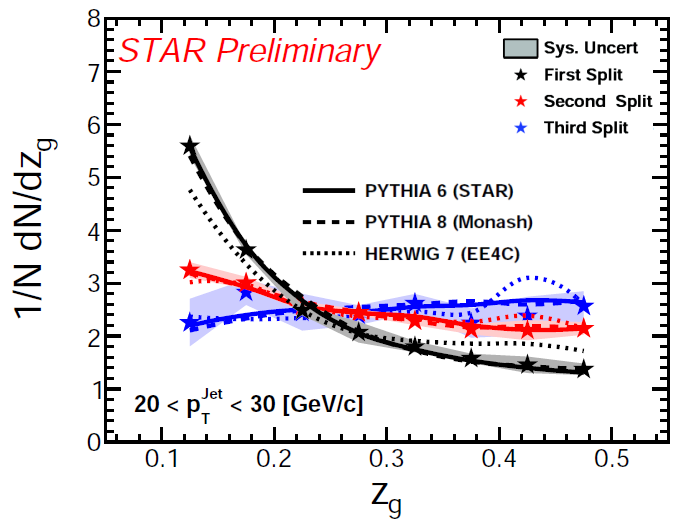}
\caption{ $z_{\text{g}}$ at the different split number for jets with $R$ = 0.4 in $p$+$p$ collisions at $\sqrt{s}$~=~200 GeV. \label{zg_rag}}
\end{figure}

Jet substructure observables show distinct correlations that are consistent with parton shower dynamics. In Fig. \ref{zg_rg_1}, the $z_{\text{g}}$ distribution steepens with wider $R_{\text{g}}$ which is consistent with the expectations from DGLAP evolution providing the appropriate perturbative description for the first hard splitting in a jet. We observe the same trend in Fig. \ref{zg_rag}, where the $z_{\text{g}}$ distribution becomes more flatter with higher split number, which means that the symmetric emission is enhanced when we go from the first to the third split. Both figures also include comparison with different Monte Carlo simulations, namely PYTHIA 6 Perugia tune \cite{PYT6, perugia}, PYTHIA 8 Monash tune \cite{PYT8, monash}, and HERWIG EE4C tune \cite{HER, ee4c}. All of the models describe the trend of the data. The main difference between HERWIG and PYTHIA lies in their parton shower and hadronization models. PYTHIA uses a $p_{\text{T}}$-ordered parton shower based on the dipole picture and employs the Lund string model for hadronization. In contrast, HERWIG uses an angular-ordered parton shower, which better preserves coherence effects, and applies a cluster hadronization model.

\begin{figure}[htb]
\includegraphics[width=0.5\linewidth]{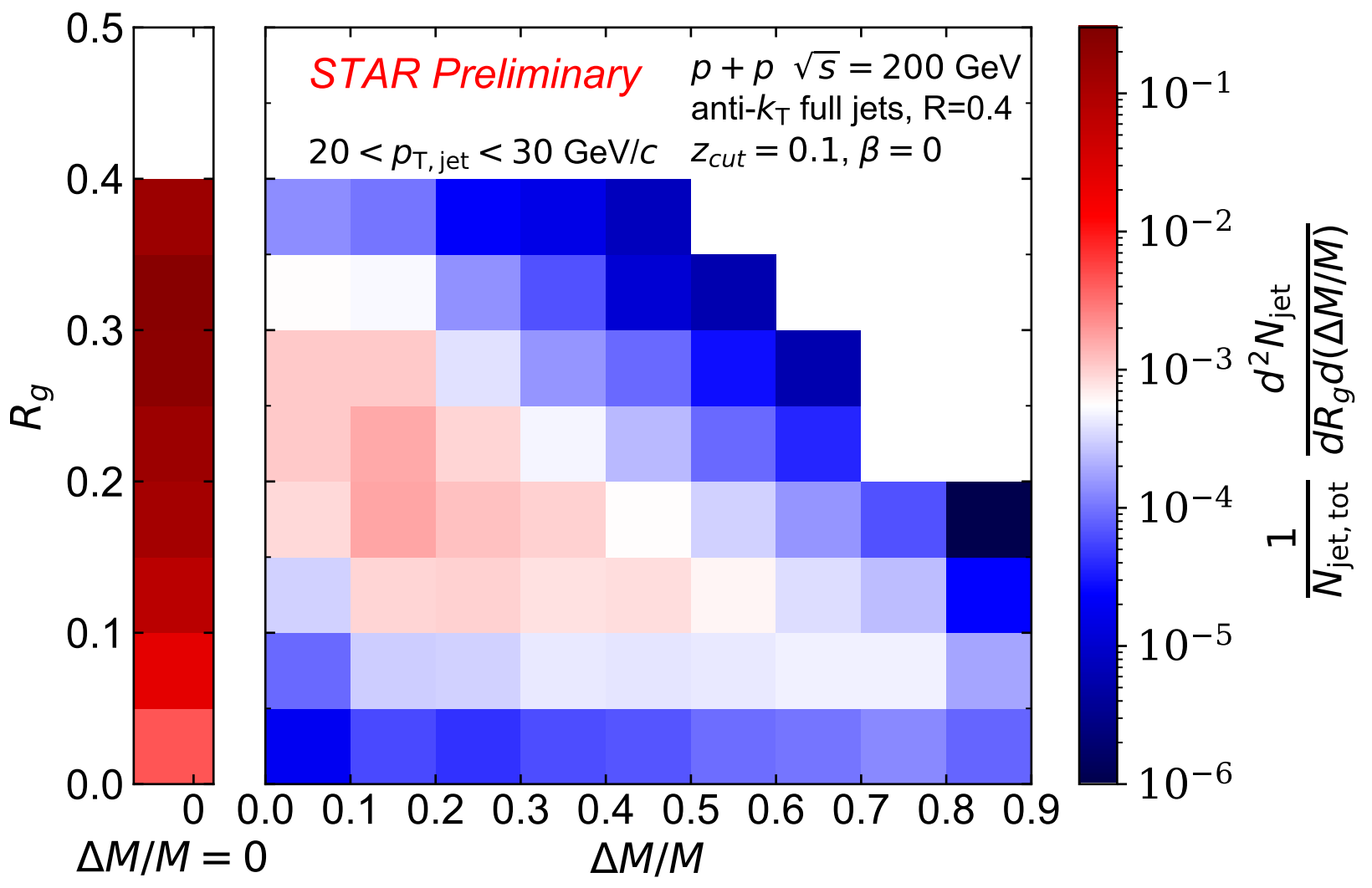}
\includegraphics[width=0.4\linewidth]{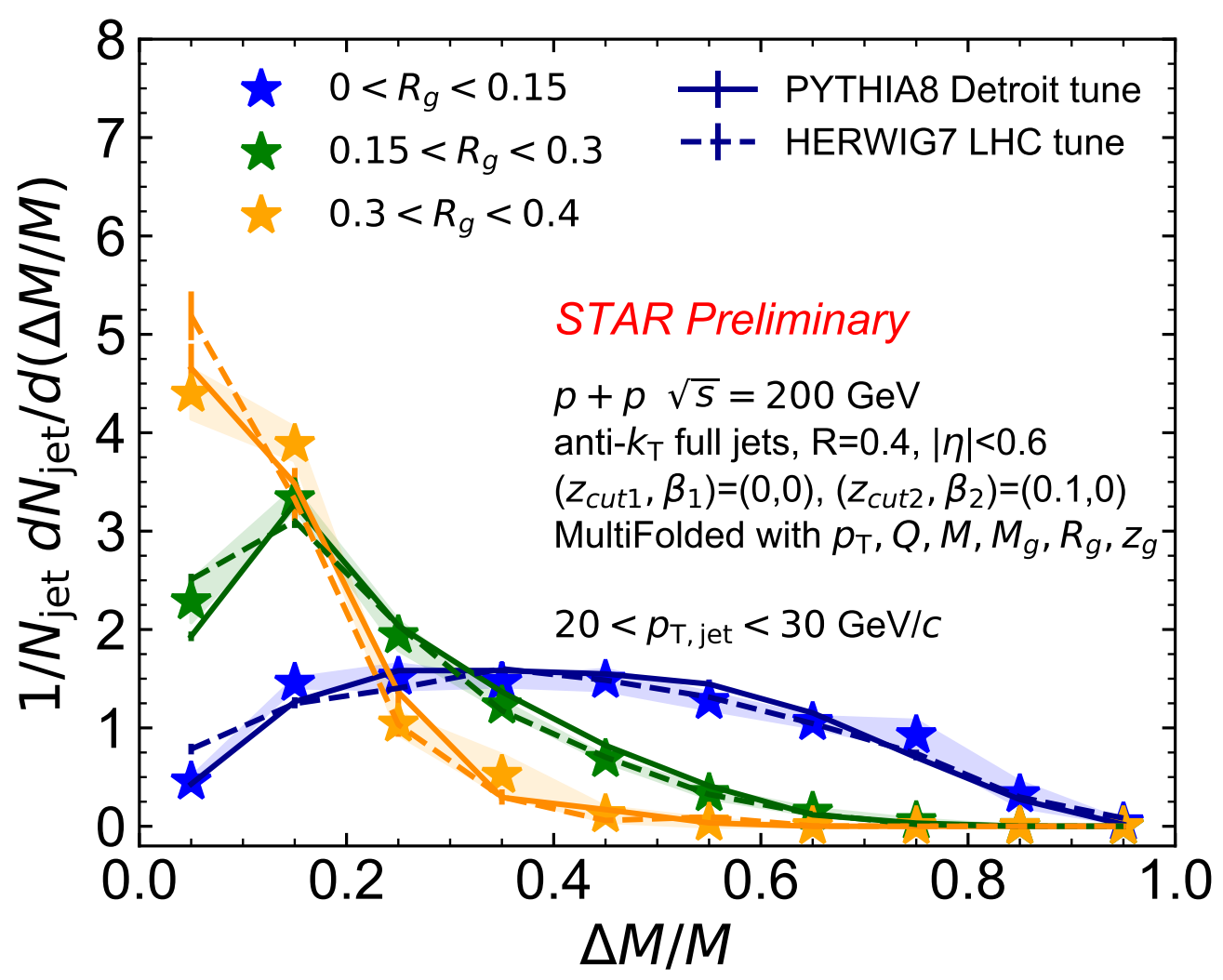}
\caption{Correlation between $\Delta M/M$ and $R_{\text{g}}$ at the first split (left) and the projection of $\Delta M/M$ for three different $R_{\text{g}}$ selections (right) for jets with $R$ = 0.4 in $p$+$p$ collisions at $\sqrt{s}$~=~200 GeV. \label{dm_rg}}
\end{figure}

In Fig. \ref{dm_rg}, $\Delta M/M$ is anti-correlated with $R_{\text{g}}$, reflecting angular ordering. We observe that for large groomed radii (indicated in yellow), the $\Delta M/M$ distribution peaks at low values, suggesting that little to no soft wide-angle radiation is present in the parton shower. The data have been compared to various Monte Carlo simulations, specifically PYTHIA 8 (Detroit tune) \cite{PYT8, Detroit} and HERWIG 7 (LHC tune), both of which reproduce the overall trend of the measurement.

\begin{figure}[htb]
\center
\includegraphics[width=0.9\linewidth]{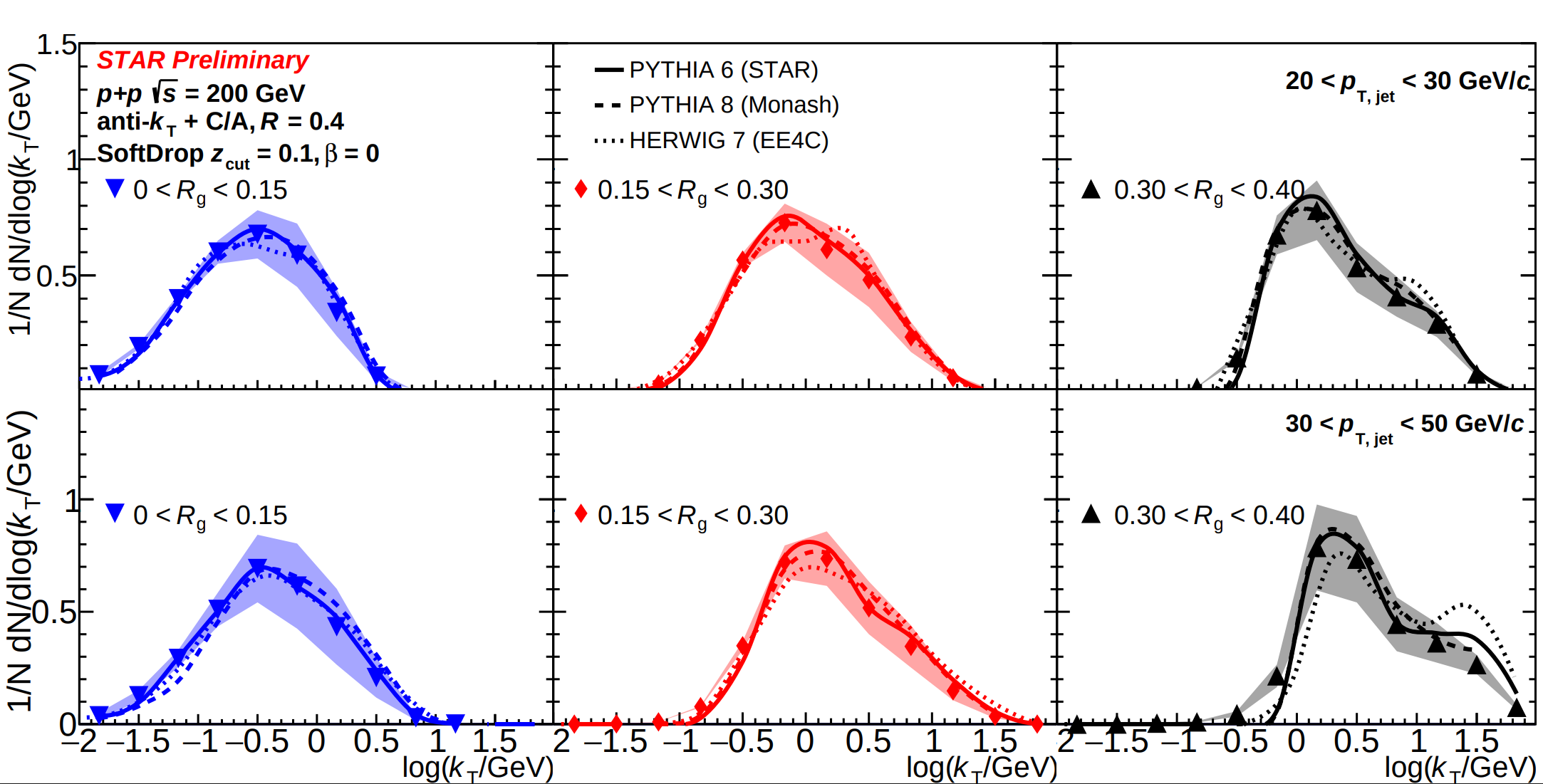}
\caption{Correlation between log($k_{\text{T}}$) and $R_{\text{g}}$ at the first split for jets with $R$ = 0.4 in $p$+$p$ collisions at $\sqrt{s}$~=~200~GeV. Individual panels correspond to different $p_{\text{T,jet}}$ intervals (see legend). \label{kT}}
\end{figure}

Figure \ref{kT} presents the distributions of log($k_{\text{T}}$) for three different selections of the groomed radius $R_{\text{g}}$ and two intervals of jet transverse momentum $p_{\text{T,jet}}$. The data clearly exhibit a strong sensitivity to variations in $R_{\text{g}}$, while the dependence on $p_{\text{T,jet}}$ remains relatively weak. Given that the zero point on the x-axis corresponds to 1 GeV, we observe a transition in the distribution from the non-perturbative regime at smaller $R_{\text{g}}$ values toward the perturbative regime at larger $R_{\text{g}}$. This behavior is consistent with the scaling of the formation time as $\tau \sim 1/R_{\text{g}}$. Overall, this measurement provides access to a wide kinematic range within the Lund Plane \cite{Lund}.

\begin{figure}[htb]
\includegraphics[width=0.45\linewidth]{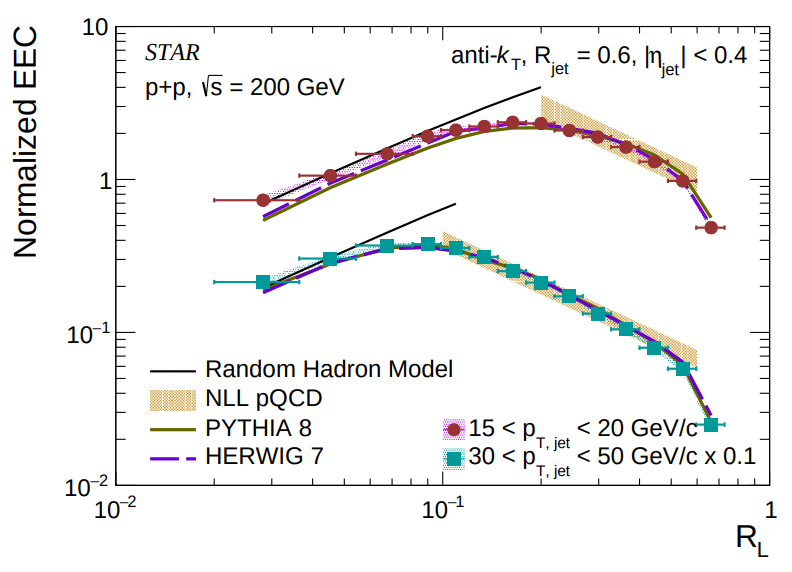}
\includegraphics[width=0.45\linewidth]{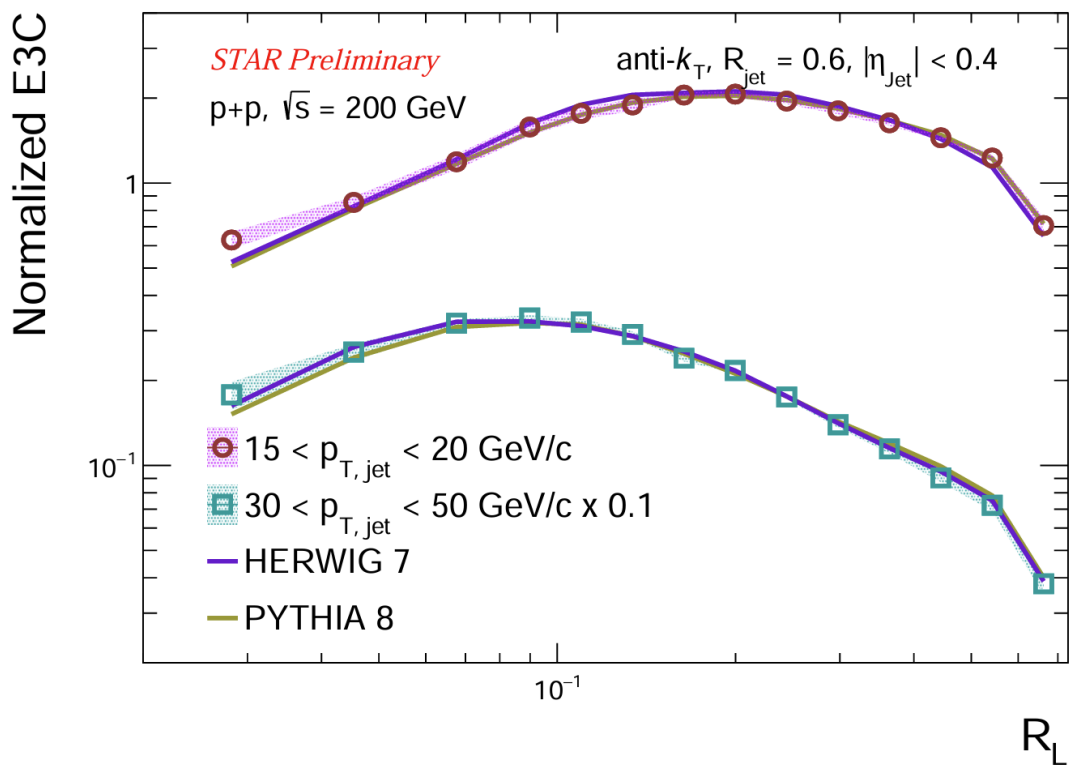}
\caption{Normalized EEC in (left) and normalized E3C (right) in $15<p_{\text{T,jet}}<20$~GeV/$c$ and $30~<~p_{\text{T,jet}}~<~50$~GeV/$c$ for jets with $R$ = 0.6 in $p$+$p$ collisions at $\sqrt{s}$~=~200~GeV. \label{EEC}}
\end{figure}

Energy correlators, EEC and E3C, in Fig. \ref{EEC} reveal a clear separation of the distribution into two distinct regimes: a non-perturbative region at low momentum transfers and a perturbative region at higher scales. These two domains are divided by a transition region, which marks the shift from non-perturbative to perturbative dynamics in the parton shower. Notably, the location of this transition is not fixed, but it systematically shifts depending on the jet transverse momentum. When the relevant observables are appropriately scaled by the jet momentum, this transition region aligns across different momentum bins, suggesting universality in the evolution of the parton shower across energy scales. The figure on the left shows the two-point energy correlator and the figure on the right shows the three-point energy correlator. The trend is the same for both EEC and E3C. Theoretical predictions and Monte Carlo models (e.g. PYTHIA 8 Detroit tune \cite{STAR_tune}, HERWIG 7) generally describe the trends of the data.

\section{Conclusion and Outlook}
In these proceedings, we present STAR preliminary results on jet substructure in $p$+$p$ collisions at $\sqrt{s}~=~200$~GeV. We presented correlations between jet substructure observables related to the first hard splitting, using the SoftDrop and CollinearDrop techniques, and complemented the analysis with energy-energy correlators. By selecting specific substructure observables and examining their correlations, including angular-dependent correlators such as the EEC, we gain access to a broad kinematic region of the Lund Plane. This approach provides valuable insights into the structure of jet showers, allowing us to disentangle perturbative dynamics from non-perturbative effects at RHIC energies. The results were compared to various Monte Carlo models, all of which qualitatively reproduce the observed trends.

\section*{Acknowledgments}
The work has been supported by the Czech Science Foundation grant 23-07499S.


\end{document}